\documentclass[12pt]{article}

\usepackage{amsmath}

\def\dspace{\baselineskip = .25in}
\def\pl{\partial}

\def\al{\alpha}
\def\bt{\beta}
\def\Ga{\Gamma}
\def\ga{\gamma}
\def\de{\delta}

\def\si{\sigma}

\def\te{\theta}

\def\La{\Lambda}
\def\lam{\lambda}

\def\om{\omega}
\def\ep{\epsilon}

\def\l{\left (}
\def\r{\right )}

\def\fr{\frac}
\def\la{\label}
\def\hs{\hspace}
\def\vs{\vspace}

\def\ran{\rangle}
\def\lan{\langle}
\def\ov{\overline}
\def\tl{\tilde}
\def\tm{\times}

\begin{document}

\dspace

\newcommand{\ba}[1]{\begin{array}{#1}} \newcommand{\ea}{\end{array}}

\numberwithin{equation}{section}


\def\Journal#1#2#3#4{{#1} {\bf #2}, #3 (#4)}

\def\NCA{\em Nuovo Cimento}
\def\NIM{\em Nucl. Instrum. Methods}
\def\NIMA{{\em Nucl. Instrum. Methods} A}
\def\NPB{{\em Nucl. Phys.} B}
\def\PLB{{\em Phys. Lett.}  B}
\def\PRL{\em Phys. Rev. Lett.}
\def\PRD{{\em Phys. Rev.} D}
\def\ZPC{{\em Z. Phys.} C}

\def\st{\scriptstyle}
\def\sst{\scriptscriptstyle}
\def\mco{\multicolumn}
\def\epp{\epsilon^{\prime}}
\def\vep{\varepsilon}
\def\ra{\rightarrow}
\def\ppg{\pi^+\pi^-\gamma}
\def\vp{{\bf p}}
\def\ko{K^0}
\def\kb{\bar{K^0}}
\def\al{\alpha}
\def\ab{\bar{\alpha}}

\def\np{Nucl. Phys. {\bf B}}\def\pl{Phys. Lett. {\bf B}}
\def\mpl{Mod. Phys. {\bf A}}\def\ijmp{Int. J. Mod. Phys. {\bf A}}
\def\cmp{Comm. Math. Phys.}\def\prd{Phys. Rev. {\bf D}}

\def\oa{\bigcirc\!\!\!\! a}
\def\ob{\bigcirc\!\!\!\! b}
\def\oc{\bigcirc\!\!\!\! c}
\def\oi{\bigcirc\!\!\!\! i}
\def\oj{\bigcirc\!\!\!\! j}
\def\ok{\bigcirc\!\!\!\! k}
\def\ve{\vec e}\def\vk{\vec k}\def\vn{\vec n}\def\vp{\vec p}
\def\vv{\vec v}\def\vx{\vec x}\def\vy{\vec y}\def\vz{\vec z}

\newcommand{\AdS}{\mathrm{AdS}}
\newcommand{\dd}{\mathrm{d}}
\newcommand{\eee}{\mathrm{e}}
\newcommand{\sgn}{\mathop{\mathrm{sgn}}}

\def\a{\alpha}
\def\b{\beta}
\def\g{\gamma}

\newcommand\lsim{\mathrel{\rlap{\lower4pt\hbox{\hskip1pt$\sim$}}
    \raise1pt\hbox{$<$}}}
\newcommand\gsim{\mathrel{\rlap{\lower4pt\hbox{\hskip1pt$\sim$}}
    \raise1pt\hbox{$>$}}}

\newcommand{\beq}{\begin{equation}}
\newcommand{\eeq}{\end{equation}}
\newcommand{\bea}{\begin{eqnarray}}
\newcommand{\eea}{\end{eqnarray}}
\newcommand{\noi}{\noindent}


\begin{flushright}
CETUP*-12/006\\
OSU-HEP-12-09
\end{flushright}

\bigskip

\begin{center}

{\Large\bf  Realistic Fermion Masses and Nucleon Decay Rates\\[0.1in] in SUSY \boldmath{$SU(5)$} with Vector--Like Matter}

\vspace{1cm}

\centerline{K.S. Babu$^{a,}$\footnote{babu@okstate.edu}, B. Bajc$^{b,c,}$\footnote{borut.bajc@ijs.si}, and Z. Tavartkiladze$^{d,}$\footnote{zurab.tavartkiladze@gmail.com}}

\vspace{0.5cm}
\centerline{$^{a}${\it\small Department of Physics, Oklahoma State University, Stillwater, OK, 74078, USA }}
\centerline{$^{b}${\it\small J.\ Stefan Institute, 1000 Ljubljana, Slovenia}}
\centerline{$^{c}${\it\small Department of Physics, University of Ljubljana, 1000 Ljubljana, Slovenia}}
\centerline{$^{d}${\it\small Center for Elementary Particle Physics, ITP, Ilia State University, 0162 Tbilisi, Georgia}}

\end{center}

\bigskip

\begin{abstract}

We show that by adding a vector--like $5+\bar{5}$ pair of matter fields to the spectrum of the
minimal renormalizable SUSY $SU(5)$ theory the wrong relations for fermion masses can be corrected,
while being predictive and consistent with proton lifetime limits.
Threshold correction from the vector--like fields improves unification of gauge
couplings compared to the minimal model.
It is found that for supersymmetric spectra lighter than $3$ TeV, which would be testable at the LHC,
at least some of the nucleon decay modes should have partial lifetimes shorter than about $2\times 10^{34}$ yrs., which is
within reach of ongoing and proposed experiments.

\end{abstract}

\thispagestyle{empty}

\clearpage


\section{Introduction}

While elegant and simple, the minimal renormalizable supersymmetric $SU(5)$  model \cite{gg,dg,sakai}
suffers from two main drawbacks. The first is the wrong predictions it makes for the light fermion masses.  This theory
predicts the asymptotic relations $m_d^0 = m_e^0$, $m_s^0 = m_\mu^0$ and $m_b^0 = m_\tau^0$ connecting
the charge $-1/3$ quark masses and charged
lepton masses, valid at the grand unification scale of $2 \times 10^{16}$ GeV. Such relations would enable one
to calculate the down--type quark masses in terms of the charged lepton masses by evolving the mass parameters via
the renormalization group equations (RGE).  The relation $m_b^0 = m_\tau^0$ is generally considered a successful prediction
of minimal SUSY $SU(5)$, since the $b$--quark mass computed in terms of $\tau$--lepton mass is typically
within about 20\% of its experimental value.  The relations involving the lighter
families, however, lead to wrong predictions.  For example, the RGE--invariant relation $m_d/m_s = m_e/m_\mu$, which follows
from the asymptotic relations of the minimal model, differs from
experimental values by about a factor of 10 ($m_d/m_s \simeq 1/20$ while $m_e/m_\mu \simeq 1/200$ at low energy scale \cite{babutasi}).

The second drawback of the minimal SUSY $SU(5)$ model is its prediction for proton lifetime for the mode
$p \rightarrow \overline{\nu} K^+$ which arises via the exchange of colored Higgsinos.  The lifetime
is generically too fast compared to the present experimental limits.
This prediction follows mainly from the requirement of gauge coupling unification.
The spectrum of the minimal supersymmetric standard model (MSSM) at low energies does not lead to a precise unification of the three
gauge couplings when the full two--loop RGE are used, and therefore requires some threshold correction from the GUT scale.
The only possibility in the minimal renormalizable $SU(5)$ set-up is to make the color triplets from the $5_H+\overline{5}_H$ Higgs fields
 (which transforms as
($3,1,-1/3) + h.c.$ under $SU(3)_C \times SU(2)_L \times U(1)_Y$ gauge group) somewhat lighter compared to the vector supermultiplets
(the $X$ and $Y$ gauge bosons of $SU(5)$).  Since the same color triplets mediate
$d=5$ proton decay \cite{Sakai:1981pk,Weinberg:1981wj},
making it lighter than the GUT scale results in a considerably shorter proton lifetime \cite{Hisano:1992jj,Lucas:1996bc,Goto:1998qg,bpw,Murayama:2001ur,bpt},
typically in conflict with experimental limits.
Notice that this outcome is due to the minimal particle content: the same color triplet that corrects the RGE running of
the gauge couplings is coupled to the Standard Model (SM) fermions with fixed Yukawa couplings. (The color triplet Yukawa couplings are
unified with the Yukawa couplings of the $SU(2)_L$ doublets also contained in $5_H+\overline{5}_H$ that generate quark and lepton masses and mixings.)
There is no other choice in the minimal model for correcting the RGE running of the gauge couplings.

There are various well known ways out of these two problems. The most commonly used  solution is the inclusion of higher dimensional
operators. Due to the vicinity of $M_{GUT}$ to $M_{\rm Planck}$ such operators may not be negligible numerically, especially for
the lighter fermion masses \cite{EmmanuelCosta:2003pu}. For example, they can easily improve the calculated masses of the first two generations. Their influence
for proton decay is even bigger. They make the Yukawa couplings to the color triplet Higgs different from those to the
weak doublet Higgs, so that there is some freedom  which can be used to somewhat suppress the $d=5$ proton decay amplitudes.
Alternatively, these higher dimensional operators can allow for a lighter color octet and weak triplet (remnants of $SU(5)$
symmetry breaking via a $24_H$) which can increase both the GUT scale and the color triplet masses \cite{Bachas:1995yt,Chkareuli:1998wi,Bajc:2002pg},
alleviating the $d=5$ proton decay problem significantly.

The problem with this natural solution is that it automatically introduces a large number of new parameters into the game, thus precluding
any quantitative prediction. So, although the model can be made consistent and realistic, it is difficult to test it. There is also some questions about
the strengths of these higher dimensional operators being of the right magnitude if they are induced by quantum gravity effects.
In this paper we take a different approach. We assume that our supersymmetric $SU(5)$ GUT is renormalizable. After all, we really do not
know how gravity influences our particle physics world, and a conservative approach would be to not rely heavily on gravity--induced
corrections. This approach of using only renormalizable couplings has brought great success in the
electroweak sector of the Standard Model. The renormalizability of the theory would greatly reduce
possible couplings in the theory resulting in enhanced predictivity.
With this in mind we shall add to the minimal supersymmetric $SU(5)$ as little as possible:
a vector-like $5+\bar 5$ matter field. This will allow unequal mixings of the down quarks and charged leptons with these fields,
thus correcting the wrong mass relations.  Simultaneously this set-up would provide a new set of color triplet/weak doublet
fields, which allows for a precise unification of gauge couplings by choosing the color triplet somewhat lighter than the
weak doublet. Note that such a choice does not run afoul with $d=5$ proton decay rates, unlike the minimal SUSY $SU(5)$ model,
since the $5+\bar{5}$ fields do not acquire vacuum expectation values (VEVs). As in minimal SUSY $SU(5)$ we assume $R$--parity
conservation, and we take the vector--like $5+\bar{5}$ pair to be fermion--like. Had we chosen Higgs--like multiplets such as
$45+ \overline{45}$, the wrong fermion mass relations could have been corrected \cite{gj}, however in this case quantitative predictions
for proton decay would be difficult to make owing to the large number of parameters that would be introduced.
Another possible solution to the wrong mass problem of the minimal SUSY $SU(5)$ model is through supersymmetric
threshold corrections arising from soft SUSY breaking terms with a particular form, see for example Ref. \cite{DiazCruz:2000mn,Enkhbat:2009jt}.  Here we shall assume that the SUSY spectrum is such that such threshold corrections
remain small.  Yet another possibility is to utilize large Yukawa couplings involving vector-like multiplets.  This can
raise the unification scale when two--loop
RGE effects are included, which would allow for a better prediction for $\alpha_3(M_Z)$ \cite{bp,hebecker}.

We now turn to the discussion of fermion masses in presence of a $5+\bar{5}$ matter fields and show how the mixing of these
fields with the MSSM fermions corrects the wrong mass relations.  We then derive the baryon number violating effective $d=5$
superpotential and study its implications for nucleon lifetime.  The small number of new parameters that are introduced with the
addition of a $5+\bar{5}$ vector--like fermions allows
the model to be consistent with current proton lifetime limits, but at the same time we find that at least some modes should
have partial lifetime less than about $2 \times 10^{34}$ yrs.  In our analysis we assume that the GUT scale stays well below the Planck
scale (by a factor of 20 to 50) so that quantum gravity effects can be ignored, and the approximate unification of the gauge
couplings that occurs in the MSSM is not a complete accident.  For supersymmetric spectrum, we assume that all super-particles have
masses less than about 3 TeV, which would make them detectable at the LHC, while at the same time providing a solution to the gauge
hierarchy problem.

\section{Fermion masses with vector--like \boldmath{$5+\bar{5}$} matter fields}

Before discussing the modifications of the fermion mass relations with the inclusion of a $5+\bar{5}$ matter fields
in SUSY $SU(5)$, let us briefly summarize the situation in the minimal renormalizable SUSY $SU(5)$ model.

\subsection{Fermion Masses in minimal SUSY \boldmath{$SU(5)$}}

The matter fields of the model consist of
three generations in representations $10_i+\bar 5_i$, $i=1,2,3$. The Higgs sector consists of an adjoint $24_H$ used
for breaking $SU(5)$ symmetry down to the SM symmetry, and a pair of $5_H +\bar{5}_H$ fields for
electroweak symmetry breaking.  The renormalizable superpotential of the adjoint field relevant
for $SU(5)$ symmetry breaking is
\beq
\label{whren}
W_{24}=\frac{m}{2}\;Tr\;(24_H^2)+\frac{\lambda}{3}\;Tr\;(24_H^3)~.
\eeq
\noi
The scalar potential induced by this superpotential has a ground state with a non-zero vacuum expectation value,
\beq
\label{gutscale}
\langle 24_H\rangle =v\;{\rm diag}\;(2,2,2,-3,3)
\eeq
\noi
which spontaneously breaks $SU(5)$ $\to$SU(3)$_C\times$SU(2)$_L\times$U(1)$_Y$.
The VEV $v$ is determined to be
\beq
v=\frac{m}{\lambda}~.
\eeq
The simplicity of Eq. (\ref{whren}) fixes the masses of the color octet (the $(8,1,0)$ fragment of $24_H$ which is a physical
Higgs particle) $M_8$ and the weak triplet (the $(1,3,0)$ fragment of $24_H$) $M_3$ to be
\beq
\label{m3jem8}
M_3=M_8=5m~.
\eeq
The same VEV sets the super-heavy $SU(5)$ gauge boson masses to be
\beq
M_X=M_Y =5\sqrt{2}g\frac{m}{\lambda}~.
\eeq

The two MSSM Higgs doublets $H_u$ and $H_d$ live in the pair of Higgs fundamentals $5_H+\bar 5_H$ and
have Yukawa couplings with the matter fields given by
\beq
\label{wyren}
W_Y=10_iY_{10}^{ij}10_j5_H+\bar 5_iY_5^{ij}10_j\bar 5_H~.
\eeq
The equality of the down--type quark masses and charged lepton masses follows from this superpotential:
\beq
M_D=\langle \bar 5_H\rangle Y_5^T=M_E^T~.
\eeq

The color triplets from $5_H + \bar{5}_H$ have the same Yukawa couplings as the Higgs doublets and would
mediate rapid proton decay via $d=5$ baryon number violating operators. For this reason they must be
ultra-heavy, preferably with a mass above the GUT scale. In the superpotential terms
\beq
W_5=\bar 5_H\left(m_H+\eta_H 24_H\right)5_H
\eeq
\noi
this can be arranged by a fine--tuning:
\beq
m_H=3\eta_H\frac{m}{\lambda}~.
\eeq
The color triplet mass is thus
\beq
M_T=5\eta_H\frac{m}{\lambda}
\eeq
\noi
which shows that $m_T$ cannot be arbitrarily large if we demand (as we do) perturbativity of the couplings:
\beq
\label{mtmx}
\frac{M_T}{M_X}=\frac{\eta_H}{\sqrt{2}g}\lsim{\cal O}(1)~.
\eeq

Due to the relation in Eq.  (\ref{m3jem8}), the requirement of gauge coupling unification would imply that
the color triplet mass is actually much lower, around or even smaller than $10^{15}$ GeV
\cite{Murayama:2001ur}.\footnote{An exception would be to choose very special MSSM
soft parameters \cite{Bajc:2002bv}. This may however require very particular
and exotic hidden and messenger sectors of SUSY breaking.} Such a light color triplet would mediate
too fast a proton decay, which is a problem with the minimal model.

\subsection{Mixing of chiral families with \boldmath{$5+ \bar{5}$} fields}

To the minimal SUSY SU(5) described in the previous subsection we now add
a vector--like pair of matter fields\footnote{The use of heavy vector-like matter to correct the bad mass
relations in GUTs is long known. For an incomplete list see for example \cite{Witten:1979nr,
Berezhiani:1985in,Davidson:1987mi,Hisano:1993uk,bb2,
Berezhiani:1995yk,Babu:1995hr,bb3,Barr:2003zx,Malinsky:2007qy,zurab,Oshimo:2009ia}.}
denoted as  $5_4+\bar 5_4$.  With their $R$--parity assumed to be identical to that of the chiral families $10_i + \bar{5}_i$
(or equivalently odd matter parity), the  most general renormalizable addition to the
superpotential of minimal $SU(5)$ is
\beq
W_4=\bar 5_a\left(\mu_a+\eta_a24_H\right)5_4,~~~ a=1,\ldots,4~.
\eeq
Notice that, without loss of generality, by an appropriate choice of the basis, the terms
$\bar 5_4 10_i \bar 5_H $ can be rotated away.
Thus, the whole Yukawa superpotential reads as
\beq
W_Y=10_iY_{10}^{ij}10_j5_H+\bar 5_iY_5^{ij}10_j\bar 5_H+\bar 5_a\left(\mu_a+\eta_a24_H\right)5_4~.
\la{extW}
\eeq
One can work in a basis where the $3\tm 3$ coupling matrix $Y_5^{ij}$ is diagonal:
$$
Y_5^{ij}=y_i\delta_{ij}~.
$$
Plugging the VEVs $\lan 5_H\ran =v_u$~, $\lan \bar 5_H\ran =v_d$~,$\lan 24_H\ran =v\;{\rm diag}\;(2,2,2,-3,-3)$ into Eq. (\ref{extW})
and keeping color triplet states $T, \bar T$ (from $5_H, \bar 5_H$), the relevant terms involving the MSSM fields and the additional vector-like
states will be
\begin{eqnarray}
W_Y &=&
L^TM_l^{4\tm 4}E^c +D^{cT}M_d^{4\tm 4}D+u^TM_U^0u^c+l^TY_5q\bar T+\fr{1}{v_u}u^TM_U^0dT \nonumber \\
&+& \, d^{cT}Y_5u^c\bar T+\fr{1}{v_u}e^{cT}M_U^0u^cT~,
\la{4-4Yuk}
\end{eqnarray}
where
$$
L^T=\l l_1, l_2, l_3, l_4\r ~,~~~~E^{cT}=\l e^c_1, e^c_2, e^c_3, \bar l_4 \r~,
$$
\beq
D^{cT}=\l d^c_1, d^c_2, d^c_3, d^c_4\r ~,~~~~D^T=(d_1, d_2, d_3, \bar d^c_4)~,~~
\la{LEc}
\eeq
\beq
\begin{array}{cc}
 & {\begin{array}{cc}
\hs{-0.6cm} &
\end{array}}\\ \vspace{1mm}
\begin{array}{c}
 \\   \\
 \end{array}\!\!\!\!\!\hs{-0.1cm}&{\!M_l^{4\tm 4}=\left(\begin{array}{cc}

 y_i\de_{ij}v_d  &~~M_i^l
\\

0& ~~|M_4^l|
\end{array}\right)}~,
\end{array}  \!\!  ~~~~~
\begin{array}{cc}
 & {\begin{array}{cc}
\hs{-0.6cm} &
\end{array}}\\ \vspace{1mm}
\begin{array}{c}
 \\   \\
 \end{array}\!\!\!\!\!\hs{-0.1cm}&{\!M_d^{4\tm 4}=\left(\begin{array}{cc}

 y_i\de_{ij}v_d & ~~M_i^d
\\
0& ~~ |M_4^d|
\end{array}\right)}~,
\end{array}  \!\!
\label{Ml-Md44}
\eeq
\beq
~M_i^l=\mu_a-3\eta_i v~,
~~~M_i^d=\mu_i+2\eta_i v~,~~~M_U^0=Y_{10}v_u~.
\label{Md-Ml-MU}
\eeq

Let us now focus on the light (MSSM) charged lepton and down--type quark masses arising from Eq. (\ref{Ml-Md44}).  These are
obtained by removing the heavy vector--like state from the spectrum.  The mass matrices of Eq. (\ref{Ml-Md44}) can be block--diagonalized
so as to bring the mass terms in the superpontential to the form
\beq
W_{mass}=e^T\hat{M}_Ee^c+d^T\hat{M}_Dd^c+u^T\hat{M}_Uu^c+M_DD\bar D+M_CC\bar C~.
\la{Wd3}
\eeq
The reduced mass matrices $\hat{M}_E$ and $\hat{M}_D$, derived in Appendix A.1, can be made real and have forms
\beq
\begin{array}{cc}
 & {\begin{array}{cc}
\hs{-0.6cm} &
\end{array}}\\ \vspace{1mm}
\begin{array}{c}
 \\   \\
 \end{array}\!\!\!\!\!\hs{-0.1cm}&{\!\hat{M_E}=\hs{-0.1cm}\left( \!\begin{array}{ccc}

 d_1c_1^e & ~0 &~0
\\
\hs{-0.5cm} -d_1s_1^es_2^e& ~d_2c_2^e &~0
\\
\hs{-0.2cm}-d_1c_2^es_1^es_3^e& -d_2s_2^es_3^e&~d_3c_3^e
\end{array}\hs{-0.1cm}\right)}~,\!
\end{array}  \!\!
\begin{array}{cc}
 & {\begin{array}{cc}
\hs{-0.6cm} &
\end{array}}\\ \vspace{1mm}
\begin{array}{c}
 \\   \\
 \end{array}\!\!\!\!\!\hs{-0.1cm}&{\!\hat{M}_D=\hs{-0.1cm}\left(\begin{array}{ccc}

 d_1c_1^d & -d_1s_1^ds_2^d & -d_1c_2^ds_1^ds_3^d
\\
0&d_2c_2^d &-d_2s_2^ds_3^d
\\
0& 0 &d_3c_3^d
\end{array}\right)}
\end{array}  \!\!
\label{ME-MD}
\eeq
with
$$
d_i=|y_iv_d|~,~~c_i^{e,d}\equiv \cos \te_i^{e,d}~,~~s_i^{e,d}\equiv \sin \te_i^{e,d}~,~~t_i^{e,d}\equiv \tan \te_i^{e,d}~,
$$
\beq
t_1^{e,d}=\fr{|M_1^{l,d}|}{|M_4^{l,d}|}~,~~~~
t_2^{e,d}=\fr{|M_2^{l,d}|}{|M_4^{l,d}|}c_1^{e,d}~,~~~~
t_3^{e,d}=\fr{|M_3^{l,d}|}{|M_4^{l,d}|}c_1^{e,d}c_2^{e,d}~.
\la{thetas}
\eeq
Note that since  $M_i^{l}\neq M_i^{d}$, the wrong GUT scale asymptotic relation $\hat{M}_E(M_{G})=\hat{M}_D^T(M_G)$, which is problematic for
the minimal renormalizable $SU(5)$ model, is avoided here.  In Eq. (\ref{Ml-Md44}) $\hat{M}_U = M_U^0 = Y_{10}v_u$, since the up--type
quarks do not mix with any of the vector--like field.

From Eq. (\ref{ME-MD}), it follows that realizing the
mass hierarchy between different families
is possible only when the diagonal factors $d_i$ are hierarchical, $d_1 \ll d_2 \ll d_3$, in which case we can write down very simple formulas for the masses:
\begin{equation}
m_i^{e,d} \simeq d_i \cos\theta_i^{e,d}~.
\end{equation}
Thus, it is possible to fit all quark and lepton masses consistently to the observed values.  The mixing angles are related by the ratios:
\begin{equation}
\label{massangle}
\frac{m^d_i}{m^e_i} \simeq \frac{\cos\theta^d_i}{\cos\theta^e_i}~.
\end{equation}

The $3 \times 3$ light fermion mass matrices are diagonalized via bi-unitary transformations
\beq
\hat{M}_E=U_E^\dag M^E_{diag}V_E~,~~~\hat{M}_D=U_D^\dag M^D_{diag}V_D~,~~~\hat{M}_U=V_u^\dag M^U_{diag}V_u^*~,~~~
\la{diag}
\eeq
by going from the flavor to the mass eigenstate  basis:\footnote{Neutrino masses are ignored for simplicity, since they are
irrelevant for our studies. They can of course be included via the seesaw mechanism with right--handed singlet neutrinos fields
introduced.  This would have very little effects on our discussions. Another possibility would be to include
bilinear R-parity violating couplings, see for example \cite{Barbier:2004ez}.}
$$
d\to U_D^T\hat{P}d~,~~~~e\to U_E^Te~,~~~~u\to V_u^TP^{1/2}u~,~~~~\nu \to U_E^T\nu
$$
\beq
d^c\to V_D^\dag \hat{P}^*d^c~,~~~~e^c\to V_E^\dag e^c~,~~~~u^c\to V_u^T \sqrt{P^*}u^c~.
\la{basis}
\eeq
The diagonal phase matrices $P$ and $\hat{P}$ are introduced (see Appendix A.1 for details)
so that the CKM matrix can be written as
\beq
V_{CKM}=\sqrt{P^*}V_u^*U_D^T\hat{P}
\la{ckm}
\eeq
in a standard parametrization with a single phase:
\beq
\begin{array}{cccc}
 & {\begin{array}{cccc}
\hs{-0.6cm} & &~&
\end{array}}\\ \vspace{1mm}
\begin{array}{c}
 \\   \\
 \end{array}\!\!\!\!\!\hs{-0.1cm}&{\!V_{CKM}=\left(\begin{array}{ccc}

 c_{12}c_{13}& ~~s_{12}c_{13} & ~~\hat{s}_{13}^*
\\
-s_{12}c_{23}-c_{12}s_{23}\hat{s}_{13}  &~~c_{12}c_{23}-s_{12}s_{23}\hat{s}_{13}  & ~~s_{23}c_{13}
 \\
s_{12}s_{23}-c_{12}c_{23}\hat{s}_{13}& ~~-c_{12}s_{23}-s_{12}c_{23}\hat{s}_{13}   &~~ c_{23}c_{13}
\end{array}\right)}~.
\end{array}  \!\!  ~~~
\label{CKM-SM}
\eeq
The entries of Eq. (\ref{CKM-SM}) can be parameterized by four Wolfenstein parameters $\lam $, $A$, $\bar \rho$ and  $\bar \eta $ as follows:
$$
s_{12}=\lam ~,~~c_{12}=\sqrt{1-\lam^2}~,~~~s_{23}=A\lam^2~,~~~c_{23}=\sqrt{1-A^2\lam^4}
$$
\beq
\hat{s}_{13}=\fr{A\lam^3(\bar \rho +i\bar \eta )\sqrt{1-A^2\lam^4}}{\sqrt{1-\lam^2}[1-A^2\lam^4(\bar \rho +i\bar \eta )]}~,~~
s_{13}=|\hat{s}_{13}|~,~~~c_{13}=\sqrt{1-s_{13}^2}~.
\la{CKMparam}
\eeq
With the central values of these parameters taken from PDG \cite{PDG}
\beq
\lam =0.2253~,~~~A=0.808~,~~~\bar \rho =0.132 ~,~~~\bar \eta =0.341 ~
\la{CKM-values}
\eeq
we can calculate the CKM elements at $M_Z$ scale. The corresponding
CKM elements at the GUT scale are obtained from $V_{CKM}(M_Z)$
by dividing the $13$, $23$, $31$ and $32$ elements by a common RGE factor ($\simeq 1.055$ for $\tan \bt =7$),
while keeping the remaining elements intact.

As far as the charged fermion masses are concerned, their  Yukawa couplings
at the GUT scale, taken to be $M_G\approx 2\cdot 10^{16}$~GeV, for $\tan \bt =7$, are taken to be
\bea
\la{YukMG}
M^U_{diag}/v_u&=&{\rm diag}\l 5.49\cdot 10^{-6},0.00323, 1\r \lam_t(\La_G),~~~\lam_t(M_G)\simeq 0.44~,\nonumber\\
M^D_{diag}/v_d&=&{\rm diag}\l 0.000886~,~0.01646~, ~1\r \lam_b(M_G),~~~\lam_b(M_G)\simeq 0.038~,\\
M^E_{diag}/v_d&=&{\rm diag}\l 0.0002777, 0.05862,1\r \lam_{\tau }(M_G),~~~\lam_{\tau }(M_G)\simeq 0.047~~.\nonumber
\eea
These values correspond to central values of these masses at low energy scale, see for eg., Ref. \cite{xing}.
These numerical values will be used below for the study of proton decay.  We emphasize that realistic fermion
masses are obtained in this model, unlike the minimal renormalizable $SU(5)$ model.

\section{The value of $\al_3(M_Z)$}

Since in the model under study we have additional states $D, \bar D, C, \bar C$ beyond those of minimal SUSY $SU(5)$, if
their masses lie below the GUT scale ($M_G$),
the unification of three gauge couplings will be modified.
The masses of these extra states are given by
$$
M_D=\sqrt{|M_1^l|^2+|M_2^l|^2+|M_3^l|^2+|M_4^l|^2}~,
$$
\beq
M_C=\sqrt{|M_1^d|^2+|M_2^d|^2+|M_3^d|^2+|M_4^d|^2}~.
\la{DCmasses}
\eeq
Since in $M_a^l, M_a^d$ there are $SU(5)$ symmetry breaking effects (see Eq. (\ref{Md-Ml-MU})), in general
these two masses differ: $M_D\neq M_C$. We will exploit this fact for improving the value of $\al_3(M_Z)$
predicted by the demand that the three gauge couplings unify.
Assuming that $M_D\simeq M_G$ and $M_C<M_G$, we will have:
\beq
\fr{1}{\al_3(M_Z)}\simeq \fr{1}{\al_3^0(M_Z)}-\fr{9}{14\pi }\ln \fr{M_C}{M_G}~,
\la{impr-al3}
\eeq
where $\al_3^0(M_Z)$ denotes the value of the strong coupling constant one would have obtained in minimal SUSY $SU(5)$ GUT.
The second term on the right--hand side of Eq. (\ref{impr-al3}) is due to the one--loop contribution of the
extra color triplet pair from the vector--like fermions with mass $M_C<M_G$.
With the choice of super-particle spectrum inspires by supergravity (see below Eq. (\ref{sugra-par}) and Table \ref{susyspectrum} for the
spectral values we use),
and with all the GUT--scale states (besides $C, \bar C$) having masses $\simeq M_G$
one would obtain   $\al_3^0(M_Z)\simeq 0.127$. To bring this somewhat large value down
we take $\fr{M_C}{M_G}\simeq 0.061$. Using this
in Eq. (\ref{impr-al3}), we obtain $\al_3(M_Z)\simeq 0.1184$ - the central value of the experimentally determined strong coupling constant.

Note that from Eq. (\ref{impr-al3}) the ratio $\fr{M_C}{M_G}$ is determined. The value of $M_G$ should be found from
the meeting point of three gauge couplings. Because of the fact that the dependance of $M_G$ on $\al_i(M_Z)$ is exponential,
we are able to determine $M_G$, and therefore also $M_T$, only to an accuracy of about $22\%$.
This will cause  an uncertainty of about $45\%$ in the $d=5$ proton decay lifetime estimate.  Further uncertainty is caused
by the uncertainty in the ratio $r = M_8/M_X$.  The natural value of $r$ is of order one, but $r \ll 1$ cannot be excluded.
Choosing $r \ll 1$ would result in larger values of the unification scale, which we shall demand to lie at least a factor
$20-50$ below the Planck scale, so that quantum gravitational corrections to the gauge coupling evolution remain small.

\section{Effective baryon number violating operators and nucleon decay}

In studying nucleon decay, we will need to derive the relevant $d=5$ baryon number violating effective operators.
These operators are obtained by integrating out the extra vector-like matter superfields, as well as the states
$T, \bar T$ from the couplings given in Eq. (\ref{4-4Yuk}). Details of this procedure are given in Appendix A.2.
Here we present the relevant effective superpotential couplings:
\beq
W_{eff}=W_{mass}+W_L^{d=5}+W_R^{d=5},
\la{Weff}
\eeq
 where $W_{mass}$ is given in Eq. (\ref{Wd3}),
\beq
W_L^{d=5}=\fr{\ep^{abc}}{M_Tv_uv_d} (u_a^T\hat{M}_Ud_b) (\nu^T\hat{M}_EP'd_c-e^T\hat{M}_EP'u_c),
\la{Wd5L-fl}
\eeq
and
\beq
W_R^{d=5}=\fr{\ep^{abc}}{M_Tv_uv_d}( u_a^{cT}\hat{M}_UP'^*e^c) (d_b^{cT}\hat{M}_D^Tu_c^c) ~.
\la{Wd5R-fl}
\eeq
Here $a, b, c$ are color indices.  $P'$ is a phase matrix $P'={\rm diag}(e^{i\de_1}, e^{i\de_2},1)$.
$M_D$ and $M_C$ are the masses of the extra vector--like weak doublets ($D, \bar D$) and color triplets ($C, \bar C$)
respectively. Note that all these coupling are written in the flavor basis of MSSM quarks and leptons.\footnote{These states differ
from those of initial superpotential (\ref{4-4Yuk}) due to various rotations (discussed in the Appendix). However, in Eqs. (\ref{Weff})-(\ref{Wd5R-fl}) we use the same notation (without primes) for simplicity.}
The couplings given in (\ref{Weff})-(\ref{Wd5R-fl}) will be needed for the discussion of
nucleon decay. Now we turn to the estimate of $d=5$ proton decay rates.

\subsection{Effective \boldmath{$d=5$} operators in the mass eigenstate basis}

With the basis change given in Eq. (\ref{basis}) and using Eqs. (\ref{diag}), (\ref{ckm}), the baryon number violating operators
of Eqs. (\ref{Wd5L-fl}), (\ref{Wd5R-fl}) will have the following form in the mass eigenstate basis:
\beq
W_L^{d=5}=\fr{\ep^{abc}}{M_Tv_uv_d}\l u_a^TPM^U_{diag}
V_{CKM}d_b\r
\left( \nu^TM^E_{diag}Vd_c-e^TM^E_{diag}VV_{CKM}^\dag u_c \right)
\la{Wd5-LLLL}
\eeq
\beq
W_R^{d=5}=\fr{\ep^{abc}}{M_Tv_uv_d}\l u_a^{cT}M^U_{diag}V_{CKM}V^\dag e^c\r
\l d_b^{cT}M^D_{diag}V_{CKM}^\dag P^*u_c^c\r ~.
\la{Wd5-RRRR}
\eeq
The matrices $V$ and $P$ are given in Eqs. (\ref{P})-(\ref{hatVform}).


The $d=6$ four fermion operator obtained from $W_L^{d=5}$ by wino dressing and involving the neutrino
has the form

\beq
{\cal O}^{d=6}_{\nu L}=\fr{\ep^{abc}}{M_T}{\cal C}^{\nu }_{\de \al \ga \rho }\l u_a^{\de }d_b^{\al }\r \l d_c^{\ga }\nu^{\rho }\r ~,
\la{d6nu}
\eeq
where
$$
{\cal C}^{\nu }_{\de \al \ga \rho }=g_2^2\sum_{\bt,\si}\left.(c_{\bt \si \ga \rho}-c_{\bt \ga \si \rho })\right|_{\mu =M_G}
\l V_{CKM}\r_{\bt \al }\l V_{CKM}^*\r_{\de \si }I(\tl{u}^{\bt},\tl{d}^\si,\tl{W})\bar A_S(d^{\ga },u^{\bt },d^{\si })
$$
$$
+g_2^2\sum_{\bt }\left.(\bar c_{\de \al \bt \rho}-\bar c_{\bt \al \de \rho })\right|_{\mu =M_G}
\l V_{CKM}\r_{\bt \ga }I(\tl{u}^{\bt},\tl{e}^{\rho },\tl{W})\bar A_S(d^{\al },u^{\de },u^{\bt })~,
$$
$$
{\rm with}~,~~~~~~c_{\bt \si \ga \rho }=\fr{1}{v_uv_d}
\l M^U_{diag}PV_{CKM}\r_{\bt \si }\l V^TM^D_{diag}\r_{\ga \rho }~,
$$
\beq
~~~~~~~~~~~~~~~~~~~~~~\bar c_{\de \al \bt \rho }=\fr{1}{v_uv_d}
\l M^U_{diag}PV_{CKM}\r_{\de \al }\l V_{CKM}^*V^TM^E_{diag}\r_{\bt \rho }~.
\la{nuC}
\eeq
Here $I$ is the loop integral defined as
\beq
I(i,j,k)=\fr{1}{16\pi^2}\fr{m_k}{m_i^2-m_j^2}\l \fr{m_i^2}{m_i^2-m_k^2}\ln \fr{m_i^2}{m_k^2}-
\fr{m_j^2}{m_j^2-m_k^2}\ln \fr{m_j^2}{m_k^2}\r~,
\la{loopI}
\eeq
while $\bar A_S$ accounts for short distance renormalization factor of the corresponding $LLLL$ d=5 operator.
Here we present some of these RG factors, which will be needed later on for numerical calculations:
$$
\bar A_S(d^{\ga },u^{\bt },d^{\si })_{\ga,\bt,\si \neq 3}=\bar A_S(d^{\al },u^{\de },u^{\bt })_{\al,\de,\bt \neq 3}\simeq 6.88~,
$$
$$
\bar A_S(d^{\ga },u^{\bt },b)_{\ga,\bt\neq 3}=\bar A_S(d^{\ga },t,d^{\si })_{\ga,\si \neq 3}=
\bar A_S(d^{\al },u^{\de },t)_{\al,\de \neq 3}\simeq 6.54~,~~
$$
\beq
\bar A_S(d^{\ga },t,b)_{\ga \neq 3}\simeq 6.2~.
\la{AS-LLLL}
\eeq
These expressions are valid for low to moderate values of $\tan \bt$.

The $d=6$ four fermion operator obtained from $W_R^{d=5}$ by  higgsino dressing and involving the neutrino has the form
\beq
{\cal O}^{d=6}_{\nu R}=\fr{\ep^{abc}}{M_T}{\cal R}^{\nu }_{\de \al \ga \rho }\l \ov{u^c}_a^{\de }\ov{d^c}_b^{\al }\r \l d_c^{\ga }\nu^{\rho }\r ~,
\la{d6nuR}
\eeq

\noindent
where

$$
{\cal R}^{\nu }_{\de \al \ga \rho }=\fr{1}{v_uv_d}\sum_{\si }\left.(\ov{\om }^*_{\de \rho \al \si }-\ov{\om }^*_{\si \rho \al \de })\right|_{\mu =M_G}
\l M^U_{diag}V_{CKM}\r_{\si \ga }\l M^E_{diag}\r_{\rho }I(\tl{e^c}^{\de},\tl{u^c}^{\si },\tl{H}^{\pm})
\bar A_{S,R}(u^{c \de}, u^{c \si })~,
$$
\beq
{\rm with}~~~\ov{\om }_{\de \rho \al \si }=\fr{1}{v_uv_d}\l M^U_{diag}V_{CKM}V^\dag \r_{\de \rho }
\l M^D_{diag}V_{CKM}^\dag P^* \r_{\al \si }~.
\la{nuR}
\eeq
 $\bar A_{S,R}$ accounts for short distance renormalization factor of the corresponding $RRRR$ d=5 operator.
 Here we give values of those, which will be needed for further calculations:
 \beq
 \bar A_{S,R}(u^c,u^{c \si })_{\si \neq 3}\simeq 4.44~,~~~~\bar A_{S,R}(u^c,t^c)\simeq 4.0~.
 \la{ASR-RRRR}
 \eeq
 
 \subsection{Nucleon decay}

The operators  responsible for $p\to \ov{\nu}_{\rho }K^+$ decay are
\beq
\fr{\ep^{abc}}{M_T}\left [{\cal C}^{\nu }_{112\rho }(u_ad_b)(s_c\nu_{\rho })+ {\cal C}^{\nu }_{121\rho }(u_as_b)(d_c\nu_{\rho })+
{\cal R}^{\nu }_{112\rho }(\ov{u^c}_a\ov{d^c}_b)(s_c\nu_{\rho })+ {\cal R}^{\nu }_{121\rho }(\ov{u^c}_a\ov{s^c}_b)(d_c\nu_{\rho })\right ]~.
\la{Op-p-Knu-rho}
\eeq

From these expressions we can calculate the partial widths for nucleon decay:
$$
\Ga (p\to \ov{\nu}_{\rho }K^+)=\fr{(m_p^2-m_K^2)^2}{32\pi m_p^3f_{\pi }^2}\left |\fr{R_L}{M_T}
\left \{ (\bt_H{\cal C}^{\nu }_{121\rho }+\al_H{\cal R}^{\nu }_{121\rho })\fr{2m_p}{3m_B}D+\right. \right.
$$
\beq
\left. \left. (\bt_H{\cal C}^{\nu }_{112\rho }+\al_H{\cal R}^{\nu }_{112\rho })\l 1+\fr{m_p}{3m_B}(D+3F)\r \right \}\right |^2~.
\la{part-nu-width}
\eeq
Here $\al_H, \bt_H$ are hadronic matrix elements and at $\mu =2$~GeV scale are \cite{Aoki:2006ib}
$|\al_H|\simeq |\bt_H|\simeq 0.012$~GeV$^3$, while the values of other parameters are $m_p=0.94$~GeV, $m_K=0.494$~GeV,
 $f_{\pi }=0.131$~GeV, $m_B=1.15$~GeV, $D=0.8, F=0.47$.
The factor $R_L \simeq 1.25$ is a long distance renormalization factor.

Note that, different from the minimal SUSY $SU(5)$ model, in Eqs. (\ref{Wd5-LLLL}) and (\ref{Wd5-RRRR}) the unitary matrix $V$ appears. This
matrix, by proper selection of its mixing angles, allows us to suppress proton decay so as to bring the partial lifetime within experimental limits.
Before demonstrating this with numerical results, in order to get a better feeling, we present an analytic study to leading order
in certain small parameters. To leading order, let us ignore (i.e., set to zero) the $2-3$ and the $1-3$ mixing angles in
the CKM matrix and in the $\hat{V}$ matrix. Let us also take the limit
$m_u, m_d, m_e\to 0$. In this limit, we get
\begin{equation}
{\cal C}^{\nu }_{1211 }={\cal C}^{\nu }_{1213}={\cal C}^{\nu }_{1121 }={\cal C}^{\nu }_{1123 }=0~.
\end{equation}
Similar results hold for the corresponding ${\cal R}^{\nu }$ amplitudes.
Therefore
\begin{equation}
\Ga (p\to \ov{\nu}_{e}K^+)=\Ga (p\to \ov{\nu}_{\tau }K^+)=0~.
\end{equation}
Only $\Ga (p\to \ov{\nu}_{\mu }K^+)$ will be non--zero due to the non--zero elements ${\cal C}^{\nu }_{1212}$ and
${\cal C}^{\nu }_{1122}$\footnote{The elements ${\cal R}^{\nu }_{1212}$, ${\cal R}^{\nu }_{1122}$ are
 suppressed strongly and can be ignored.}
which are given by
\beq
{\cal C}^{\nu }_{1212}= {\cal C}^{\nu }_{1122}\simeq g_2^2\l I(\tl{u}, \tl{d})+I(\tl{u}, \tl{e})\r \bar A_S^{\al }e^{i\om_2}\lam_s\lam_{\mu }\sin \te_c
\l \sin \te_ce^{i(\phi_2+\de_2)}+\hat{V}_{21}e^{i\phi_1}\r ~.
\la{Cs}
\eeq
Note that in the limit $\hat{V}_{21}\to 0$ the expressions of Eq. (\ref{Cs}) will coincide with those of minimal SUSY $SU(5)$.
Now, we can select the matrix element $\hat{V}_{21}$ in such a way that these coefficients vanish (or are suppressed):
$\sin \te_ce^{i(\phi_2+\de_2)}+\hat{V}_{21}e^{i\phi_1}=0$, or
\beq
|\hat{V}_{21}|=\sin \te_c ~,~~~~{\rm Arg}(\hat{V}_{21})=\pi +\phi_2+\de_2-\phi_1 ~.
\la{selectV21}
\eeq
With this conditions satisfied we get $\Ga (p\to \ov{\nu}_{\mu }K^+)\simeq 0$ and the decay $p\to \ov{\nu}K^+$ will be eliminated.
Note that the conditions in Eq. (\ref{selectV21}) are easily satisfied. This is true for the second relation because all phases entering there are free.
As far as the condition  $|\hat{V}_{21}|=\sin \te_c$ is concerned, from (\ref{hatVform}), with $t_1^es_2^e \stackrel{<}{_\sim }5t_1^ds_2^d$ we have
$|\hat{V}_{21}|\approx \fr{m_d}{m_s}t_1^ds_2^d$. With the selection $t_1^ds_2^d\approx 4$ we get $|\hat{V}_{21}|\approx 0.2\approx \sin \te_c$.

With the inclusion of $1-3$ and $2-3$ mixings, and $m_{u,d,e}\neq 0$, the expressions get more lengthy, making analytical treatment harder.
Thus, in the following we proceed with a numerical study, demonstrating the possibility of proton lifetime suppression.

\subsection{Exact numerical results}

Following Eq. (\ref{massangle}) we choose
\beq
\te_1^l=\arccos \l \fr{m_e}{m_d}\cos \te_1^d\r~,~~~~\te_2^d=\arccos \l \fr{m_s}{m_{\mu}}\cos \te_2^l\r~,
\te_3^d=\arccos \l \fr{m_b}{m_{\tau}}\cos \te_3^l\r~.
\la{relations}
\eeq
Then there are only three independent angles. We treat $\te_1^d$, $\te_2^l$ and $\te_3^l$ as free parameters and select
them in such a way as to suppress $d=5$
proton decay rates adequately. We also have the free phases $\de_{1,2}, \om_{1,2}, \phi_{1,2}$, which we vary so as to suppress proton decay rate.

For soft SUSY breaking parameters we adopt supergravity--inspired spectrum.
However, we deviate from mSUGRA and allow for non-universality in the Higgs
boson mass. This is implemented by taking the pseudoscalar Higgs mass $M_A$
and $\mu$ as independent parameters.  At the GUT scale we take as input, inspired
by the ``natural SUSY" spectrum of Ref. \cite{Baer:2012up},
\bea
M_0=3\;{\rm TeV},&~&M_{1/2}=568.3\;{\rm GeV},~~~A_0=-5\;{\rm TeV},\nonumber \\
\tan{\beta}=7,&~&\hskip 0.4cm\hskip 0.4cm
\mu=150~{\rm GeV},~~~M_A=1~{\rm TeV},
\la{sugra-par}
\eea
where $M_0$ ($M_{1/2}$) is the usual universal soft mass for chiral matter superfields
(gauginos) at the GUT scale, $A_0$ the common trilinear term, while the Higgs sector
is not universal ($M_{H_{u,d}}^2\ne M_0^2$). The value of $\tan{\beta}$ given is at the weak scale,
corresponding to $\tan{\beta}=6.75$ at the GUT scale.
The parameters are chosen so that the SUSY spectrum is lighter than approximately 3 TeV, which can be discovered at LHC.
For numerical calculations we used the code SuSpect \cite{Djouadi:2002ze}, through which we make sure that the lightest
(SM like) Higgs mass is $\simeq 125$ GeV.
The  spectrum (at weak scale) we get for the input of Eq. (\ref{sugra-par}) is given in Table \ref{susyspectrum}.
\begin{table}[t]
\hskip 1.6cm \begin{tabular}{|c|c|c|c|c|c|c|c|c|c|c|}
\hline
  h & A & $H^0$ & $H^\pm$ & $\tilde \chi^\pm_1$ & $\tilde \chi^\pm_2$
  & $\tilde \chi^0_1$ & $\tilde \chi^0_2$ & $\tilde \chi^0_3$
  & $\tilde \chi^0_4$ & $\tilde g$  \\
\hline
  125 & 1000 & 1000 & 1003 & 145 & 497 & 132 & -158 & 259 & 497 & 1450  \\
\hline
\end{tabular} \par
\hskip 2.6cm \begin{tabular}{|c|c|c|c|c|c|c|c|}
\hline
  $\tilde t_1 $ & $\tilde t_2$ & $\tilde u_1, \tilde c_1$ & $\tilde u_2, \tilde c_2$
  & $\tilde b_1$ & $\tilde b_2$ & $\tilde d_1, \tilde s_1$ & $\tilde d_2, \tilde s_2$  \\
\hline
  554 & 2197 & 3144 & 3241 & 2186 & 3096 & 3145 & 3118  \\
\hline
\end{tabular} \par
\hskip 3.7cm \begin{tabular}{|c|c|c|c|c|c|}
\hline
  $\tilde \tau_1$ & $\tilde \tau_2$ & $\tilde e_1, \tilde \mu_1$ & $\tilde e_2, \tilde \mu_2$
  & $\tilde \nu_\tau$ & $\tilde \nu_e, \tilde \nu_\mu$  \\
\hline
  2849 & 3062 & 3073 & 2871 & 3061 & 3072  \\
\hline
\end{tabular}
\caption{Particle masses (in GeV) obtained  by the input given in Eq. (\ref{sugra-par}) in MSSM.}
\label{susyspectrum}
\end{table}
These values will be used in the calculation of proton lifetime.

One choice of the three free angles and phases giving adequate suppression of proton decay rate is:
$$
\te_1^d=1.3433 ,~~~~~~\te_2^l=1.016 ,~~~~~~\te_3^l=0.10275,~~
$$
$$
\phi_1=\de_1=0,~~~~\phi_2=3.3065,~~~~~\de_2=1.883,
$$
\beq
\om_1=2.515~,~~~~~~~~~~\om_2=1.748.
\la{par-select}
\eeq
With these input values we obtain for the decay rate $p \rightarrow \overline{\nu} K^+$
$$
\Ga_{d=5}^{-1}(p\to \bar{\nu }K^+)
=\fr{1}{\sum_{i=1}^3\Ga_{d=5}(p\to \bar{\nu_i}K^+)}\simeq
$$
\beq
4\cdot 10^{33}\hs{0.1cm}{\rm yrs} \tm
\l \fr{0.012{\rm GeV}^3}{\bt_H}\r^2  \l \fr{1.25}{R_L}\r^2
\l \fr{M_T}{4.8\!\cdot \!10^{16}{\rm GeV}}\r^2.
\la{tot-pdecay}
\eeq
In Table \ref{t:t1}
we summarize the partial lifetimes for this and other decay modes.
Not all decay modes (induced by the $d=5$ operators) are listed,
those with lifetimes exceeding $\sim 5\cdot 10^{36}$ years are not shown. Note that with
further tuning of parameters, we may suppress
even more the $p\to \bar{\nu }K^+$ decay. However, we can not decrease much further the value
of $M_T$ because that would decrease the lifetime $\Ga_{d=5}^{-1}(p\to \mu^+K^0)$ whose value is
already near at the experimental limit \cite{superk} (see Table \ref{t:t1}).

Note that with the value $M_T=4.8\cdot 10^{16}$~GeV (used in Eq. (\ref{tot-pdecay})), the mass of the $SU(5)$ gauge bosons $(X,Y)$ should be greater than about  $2 \times 10^{16}$~GeV in order to be consistent with
perturbtativity \cite{Hisano:1993uk}. Such
 a value for $M_X$ would mean that there is some chance for the observation of the gauge boson mediated
nucleon decay such as $p \rightarrow e^+ \pi^0$, but this will be
challenging.

One can try to increase the color triplet mass to further suppress the
rates for the $d=5$ modes.
Due to the perturbativity constraint (see Eq. (\ref{mtmx})) one needs first to increase the heavy gauge boson mass. For $m_3=m_8$ this equals
\beq
M_X=M_X^0/r^{1/3}
\eeq
\noi
where $M_X^0\approx 2.10^{16}$ GeV.
By choosing $r\approx 1/10$ or so $M_X$ and thus $M_T$ can be increased by a factor of $2$.
The color triplet mass can now be raised to $M_T\approx 10^{17}$ GeV, which would imply the scaling of all lifetimes for all modes in Table \ref{pdkmodes} upward
by a factor of 4. Further increase of the triplet mass could jeopardize the expansion in
inverse powers of the Planck scale, so we will not consider it.  We see that, with the assumption that SUSY particles masses
lie below about 3 TeV, which is testable at the LHC, proton lifetime cannot exceed about $2 \times 10^{34}$ years.
This is within reach of ongoing and proposed experiments.

%
%
\begin{table}[h]
\caption{\label{pdkmodes} Inverse widths for nucleon decay. Calculations are carried out
for the SUSY parameters (spectrum) given in Eq. (\ref{sugra-par}), Table \ref{susyspectrum}. The model
parameters are given in Eqs. (\ref{par-select}), (\ref{relations}), along with $M_T = 4.8 \times 10^{16}$ GeV. Other parameters used can be found right after Eq. (\ref{part-nu-width}).
 }

\label{t:t1} $$\begin{array}{|c||c|}

\hline
\vs{-0.3cm}
 & \\

\vs{-0.4cm}

\Ga_{d=5}^{-1}(p\to \bar{\nu}K^+)&4 \cdot 10^{33}~{\rm yrs}.\\

& \\

\hline

\vs{-0.3cm}
 &  \\

\vs{-0.3cm}
\Ga_{d=5}^{-1}(n\to \bar{\nu}K^0)& 2 \cdot 10^{33}~{\rm yrs}.\\

&  \\

\hline

\vs{-0.3cm}
 &  \\

\vs{-0.3cm}
\Ga_{d=5}^{-1}(p\to \mu^+K^0)& 1.0\cdot 10^{34}~{\rm yrs}.\\

&  \\

\hline

\vs{-0.3cm}
 &  \\

\vs{-0.3cm}
\Ga_{d=5}^{-1}(p\to \mu^+\pi^0)& 1.8 \cdot 10^{34}~{\rm yrs}.\\

&  \\

\hline

\vs{-0.3cm}
 &  \\

\vs{-0.3cm}
\Ga_{d=5}^{-1}(p\to \bar{\nu}\pi^+)&7.3 \cdot 10^{33}~{\rm yrs}. \\

&  \\

\hline

\vs{-0.3cm}
 &  \\

\vs{-0.3cm}
\Ga_{d=5}^{-1}(n\to \bar{\nu}\pi^0)& 1.5 \cdot 10^{34}~{\rm yrs}.\\

&  \\

\hline









\end{array}$$

\end{table}
%
%

We have not included gluino dressing of the effective $d=5$ operators in order to obtain four fermion operators
for proton decay.  When universality is assumed, as we do, for the masses of the superpartners of the chiral fermions,
the gluino dressing diagrams are highly suppressed \cite{ellis} compared to the Wino dressing diagrams.  This is primarily due to
the antisymmetric nature of the $QQQL$ operator in flavor.  With the SUSY particle masses taken to be less than about
3 TeV, universality in the soft scalar masses is almost a necessity in order to suppress flavor changing neutral currents
(FCNC) arising from the exchange of SUSY particles.  If the third family squark and slepton masses are taken to be different from those
of the (degenerate) first two families, FCNC processes may not be excessive.  In this case, the gluino dressing contributions
to nucleon decay may become important, but typically the amplitude is not much more than that arising from the Wino dressing,
see for eg. discussions in Ref. \cite{bb4}.  Thus, variation of SUSY spectrum would not significantly alter the upper limit
on nucleon lifetime derived above, as long as the sparticle masses lie below 3 TeV or so.

\section{Conclusions}

In this paper we have shown that the main problems of the minimal renormalizable model based on SUSY $SU(5)$ can
be cured by adding a vector--like pair of $5+\bar{5}$ matter fields.  This allows for the mixing of chiral families
with the vector--like fields, which we show corrects the wrong mass relations of minimal $SU(5)$.  The mass splitting
between the color triplets and the weak doublets of this vector--like fields improves the unification of the three
gauge couplings.  The color triplets from the $5_H + \bar{5}_H$ fields, which mediate $d=5$ proton decay can have
GUT scale masses, thus avoiding the rapid proton decay problem of the minimal model.  The small number of couplings of this model enables us to make quantitative predictions for partial lifetimes for proton decay.  We find that, in the favorable case that the LHC is sensitive to the
discovery of the whole SUSY spectrum (corresponding to all the super-partner masses and Higgs boson masses $\lsim 3$ TeV), at least some of the modes
should have partial lifetimes shorter than about $2\times 10^{34}$ yrs, which is within reach of proposed experiments.

\section*{Acknowledgments}
This work has been supported in part by the Slovenian Research Agency (BB), and by the Slovenia-USA
program BI-US/09-12-036 (KSB and BB). BB thanks the Physics Department of the Oklahoma State University
in Stillwater, Oklahoma, for hospitality.  The work of KSB is supported in part by DOE Grant No. DE-FG02-04ER41306.
ZT thanks Shota Rustaveli National Science Foundation for partial support  (contract \#03/79).
We thank the Center for Theoretical Underground Physics and Related Areas, CETUP* 2012, Lead, South Dakota, where this work was completed for its hospitality and for partial support.


\appendix

\section{Deriving $W_{eff}$}

In this Appendix, we give details of obtaining the effective superpotential, both for the
light fermion mass matrices, and for the $d=5$ baryon number violating superpotential couplings.
The effective superpotential is obtained by decoupling the extra heavy vector like states.
First we integrate out the extra matter states. This is performed by block--diagonalization of the first two
coupling matrices in Eq. (\ref{4-4Yuk}).

\subsection{Derivation of $W_{mass}$}

With the transformation
\beq
L=P_lV_l^\dag L'~,~~~E^c=P_{e^c}{E^c}'~,~~~D^c=P_{d^c}V_d^\dag {D^c}'~,~~~D=P_{q}D'~,
\la{transfs}
\eeq
the matrices $M_l^{4\tm 4}$ and $M_d^{4\tm 4}$ get transformed to \cite{bb3}

\beq
\begin{array}{cccc}
 & {\begin{array}{cccc}
\hs{-0.6cm} & &~&
\end{array}}\\ \vspace{1mm}
\begin{array}{c}
 \\   \\
 \end{array}\!\!\!\!\!\hs{-0.1cm}&{\!M_l^{4\tm 4}\to V_lP_lM_l^{4\tm 4}P_{e^c}=\left(\begin{array}{cc}

  \hat{M}_E&0
\\
{\cal O}(v_d)&M_D
\end{array}\right)}~,
\end{array}  \!\!  ~~~
\label{diagMl44}
\eeq

\beq
\begin{array}{cccc}
 & {\begin{array}{cccc}
\hs{-0.6cm} & &~&
\end{array}}\\ \vspace{1mm}
\begin{array}{c}
 \\   \\
 \end{array}\!\!\!\!\!\hs{-0.1cm}&{\!M_d^{4\tm 4}\to V_dP_{d^c}M_d^{4\tm 4}P_e=\left(\begin{array}{cccc}

  \hat{M}_D^T&0
\\
{\cal O}(v_d)&M_C
\end{array}\right)}~.
\end{array}  \!\!  ~~~
\label{diagMd44}
\eeq
The matrices in Eq. (\ref{transfs}) are given by
$$
P_l=e^{i\om_l}{\rm Diag}\l e^{-i\phi_{M_1^l}}~,~e^{-i\phi_{M_2^l}}~,~e^{-i\phi_{M_3^l}}~,~1\r
$$
$$
P_{e^c}=e^{-i\om_l}{\rm Diag}\l e^{i(\phi_{M_1^l}-\phi_{y_1v_d})}~,~e^{i(\phi_{M_2^l}-\phi_{y_2v_d})}~,~e^{i(\phi_{M_3^l}-\phi_{y_3v_d})}~,~1\r
$$
$$
P_{d^c}=e^{i\om_{d^c}}{\rm Diag}\l e^{-i\phi_{M_1^d}}~,~e^{-i\phi_{M_2^d}}~,~e^{-i\phi_{M_3^d}}~,~1\r
$$
\beq
P_q=e^{-i\om_{d^c}}{\rm Diag}\l e^{i(\phi_{M_1^d}-\phi_{y_1v_d})}~,~e^{i(\phi_{M_2^d}-\phi_{y_2v_d})}~,~e^{i(\phi_{M_3^d}-\phi_{y_3v_d})}~,~1\r
\la{44Phases}
\eeq

\beq
\begin{array}{cccc}
 & {\begin{array}{cccc}
\hs{-0.6cm} & &~&
\end{array}}\\ \vspace{1mm}
\begin{array}{c}
 \\   \\
 \end{array}\!\!\!\!\!\hs{-0.1cm}&{V_{l,d}=\left(\begin{array}{cccc}

 c_1^{e,d} & ~~0 & ~~0 &-s_1^{e,d}
\\
 -s_1^{e,d}s_2^{e,d} &~~ c_2^{e,d} & ~~ 0&-c_1^{e,d}s_2^{e,d}
 \\
-c_2^{e,d}s_1^{e,d}s_3^{e,d}& ~~ -s_2^{e,d}s_3^{e,d}  &~~c_3^{e,d} &-c_1^{e,d}c_2^{e,d}s_3^{e,d}
\\
c_2^{e,d}c_3^{e,d}s_1^{e,d}& ~~ c_3^{e,d}s_2^{e,d}  &~~ s_3^{e,d}&c_1^{e,d}c_2^{e,d}c_3^{e,d}
\end{array}\right)}~,
\end{array}  \!\!  ~~~
\label{Vld}
\eeq
where definitions for the entries of Eq. (\ref{Vld}) see Eq. (\ref{thetas}).  We use the notation
$\phi_X$ to denote the phase of a complex parameter $X$.  Thus $\phi_{y_1v_d}$ is the argument
of $y_1v_d$, etc.
With all these, one can easily check that the matrices $\hat{M}_E$, $\hat{M}_D$ and masses
$M_D, M_C$ are given by Eqs. (\ref{ME-MD}) and (\ref{DCmasses}) respectively.
The entries ${\cal O}(v_d)$ in Eqs. (\ref{diagMl44}), (\ref{diagMd44}) can be safely ignored.
Thus, the diagonal block-entries in these matrices, together with $\hat{M}_U$, coincide with the terms of Eq. (\ref{Wd3}).

\subsection{Deriving effective $d=5$ operators}

Now we turn to the derivation of the effective $d=5$ baryon number violating superpotential couplings.
With the transformations of Eq. (\ref{transfs}) and with
\beq
q=P_q'q'~,~~~~~u^c=P_q'{u^c}'~,
\la{Qtransf}
\eeq
where
\beq
P_q'=e^{-i\om_{d^c}}{\rm Diag}\l e^{i(\phi_{M_1^d}-\phi_{y_1v_d})}~,~e^{i(\phi_{M_2^d}-\phi_{y_2v_d})}~,~e^{i(\phi_{M_3^d}-\phi_{y_3v_d})}\r~,
\la{Pqpr}
\eeq
one can derive the couplings of the light states with the color triplets $T, \bar T$:

\beq
\fr{1}{v_d}l^T\hat{M}_EP'q\bar T+\fr{1}{v_u}u^T\hat{M}_UdT+
\fr{1}{v_d}d^{cT}\hat{M}_D^Tu^c\bar T+\fr{1}{v_u}e^{cT}P^{'*}\hat{M}_Uu^cT~,
\la{fl-T}
\eeq
where we have omitted primes for the quark and lepton states. The matrix $P'$,
 without loss of  generality, can be parameterized as:
\beq
P'={\rm Diag}\l e^{i\de_1},e^{i\de_2},1\r ~.
\la{Ppr}
\eeq
Further, integrating out the states $T, \bar T$ with mass $M_T$, from Eq. (\ref{fl-T}) we derive the effective
$d=5$ operators given in Eqs. (\ref{Wd5L-fl}), (\ref{Wd5R-fl}). These are written in a flavor basis.

Finally, we present the matrices which appear in the $d=5$ couplings written in the the mass eigenstate basis,
using the transformations given in Eq. (\ref{basis}). These are the phase matrix $P$
\beq
P={\rm Diag}\left(e^{i\omega_1},e^{i\omega_2},1\right)~,~~~
\la{P}
\eeq
and the matrix
\beq
V=\hat{V}\hat{P}~,~~~~~{\rm with}~~~~\hat{V}=V_EP'U_D^T~,~~~
~\hat{P}={\rm Diag}\left(e^{i\phi_1},e^{i\phi_2},1\right)~.
\la{VhatP}
\eeq
The elements of the matrix $\hat{V}$ are:
$$
\hat{V}_{11}\simeq e^{i\de_1}~,~~~~~~~~\hat{V}_{12}\simeq -\fr{m_d}{m_s}t_1^ds_2^de^{i\de_1}+\fr{m_e}{m_{\mu }}t_1^es_2^ee^{i\de_2}~,
$$
$$
\hat{V}_{13}\simeq -\fr{m_d}{m_b}t_1^dc_2^ds_3^de^{i\de_1}-\fr{m_s}{m_b}\fr{m_e}{m_{\mu }}t_1^es_2^et_2^ds_3^de^{i\de_2}+\fr{m_e}{m_{\tau}}t_1^e\fr{s_3^e}{c_2^e}~,
$$
$$
\hat{V}_{21}\simeq \fr{m_d}{m_s}t_1^ds_2^de^{i\de_2}-\fr{m_e}{m_{\mu }}t_1^es_2^ee^{i\de_1}~,~~~~~~\hat{V}_{22}\simeq e^{i\de_2}~,~~~~~
\hat{V}_{23}\simeq -\fr{m_s}{m_b}t_2^ds_3^de^{i\de_2}+\fr{m_{\mu}}{m_{\tau}}t_2^es_3^e~,
$$
$$
\hat{V}_{31}\simeq \fr{m_d}{m_b}t_1^d\fr{s_3^d}{c_2^d}-\fr{m_d}{m_s}\fr{m_{\mu}}{m_{\tau }}t_2^es_3^et_1^ds_2^de^{i\de_2}-
\fr{m_e}{m_{\tau}}t_1^es_3^ec_2^ee^{i\de_1}~,
$$
\beq
\hat{V}_{32}\simeq \fr{m_s}{m_b}t_2^ds_3^d-\fr{m_{\mu}}{m_{\tau}}t_2^es_3^ee^{i\de_2}~,~~~~~~~\hat{V}_{33}\simeq 1~.
\la{hatVform}
\eeq

\subsection{An alternative derivation of \boldmath{$W_{mass}$}}

Here we provide an alternative, perhaps more intuitive, derivation of the effective
mass matrices for the down--type quarks and charged leptons that follow from Eq. (\ref{Ml-Md44}).
We write down these matrices in a unified $SU(5)$ notation,
\beq
\label{initial}
{\cal L}=
\begin{pmatrix}
\bar 5_i & \bar 5_4
\end{pmatrix}
\begin{pmatrix}
m^0_{ij} & M_i \cr 0 & M_4
\end{pmatrix}
\begin{pmatrix}
10_j \cr 5_4
\end{pmatrix}
\eeq
where
\bea
m^0_{ij}&=&y_i\delta_{ij}\langle \bar 5_H\rangle \\
M_a&=&\mu_a+\eta_a\langle 24_H\rangle
\hskip 0.5cm , \hskip 0.5cm a=1\ldots 4
\eea
Here $\langle 24_H\rangle = 2v$ for the color triplet quark fields, while $\langle 24_H\rangle = -3v$ for
the $SU(2)_L$ doublet lepton fields from the $\overline{5}_a + 5_4$.  Now we make a unitary rotation parametrized by
\beq
\begin{pmatrix}
\bar 5_i & \bar 5_4
\end{pmatrix}
\to
\begin{pmatrix}
\bar 5_i & \bar 5_4
\end{pmatrix}
U
\label{rot}
\eeq
with
\bea
\label{unitary}
U&=&
\begin{pmatrix}
\Lambda & -\Lambda x \cr x^\dagger \bar\Lambda & \bar\Lambda
\end{pmatrix}\\
x^T&=&(M_1,M_2,M_3)/M_4\\
\label{Lambda}
\Lambda&=&(1+x\,x^\dagger)^{-1/2}\hskip 0.5cm ,
\hskip 0.5cm \bar\Lambda=(1+x^\dagger\,x)^{-1/2}=(1+|x|^2)^{-1/2}
\eea
Note that the unitary matrix $U$ is different for the quarks and leptons, since the $M_i$ factors
that enter into $U$ are different.  Similarly, the $x_i$ factors are not the same in these
two sectors.  We shall not explicitly show here the dependence of
$U$ or $x_i$ on the fermion flavor, but it is to be understood.

With the rotation of Eq. (\ref{rot}) , Eq. (\ref{initial}) becomes
\beq
{\cal L}\to
\begin{pmatrix}
\bar 5_i & \bar 5_4
\end{pmatrix}
\begin{pmatrix}
(\Lambda m^0)_{ij} & 0 \cr (x^\dagger\,\bar\Lambda\,m^0)_i &
x^\dagger\,\bar\Lambda\,M+\bar\Lambda\,M_4
\end{pmatrix}
\begin{pmatrix}
10_j \cr 5_4
\end{pmatrix}
\eeq
The heavy pair is now $\bar 5_4-5_4$, and the light mass matrices for down quarks and
charged leptons become
\beq
M^D=\Lambda^d m^0\hskip 0.5cm\,\hskip 0.5cm M^E=m^0\Lambda^{eT}
\label{lightmass}
\eeq
with
\beq
x_{Di}=\frac{\mu_i+2\eta_i v}{\mu_4+2\eta_4 v}\hskip 0.5cm,\hskip 0.5cm
x_{Ei}=\frac{\mu_i-3\eta_i v}{\mu_4-3\eta_4 v}
\eeq
where we have explicitly shown the separate matrices for down type quarks and charged leptons, using the GUT scale VEV $v$
given in Eq. (\ref{gutscale}).

The matrix $\Lambda$ from (\ref{Lambda}) (for each sector separately) can be written explicitly as
\bea
\Lambda&=&1-\frac{x\,x^\dagger}{\sqrt{1+|x|^2}\left(\sqrt{1+|x|^2}+1\right)}
\nonumber\\
&=&
\begin{pmatrix}
1-c\,x_1\,x_1^* & -c\,x_1\,x_2^* & -c\,x_1\,x_3^* \cr
-c\,x_2\,x_1^* & 1-c\,x_2\,x_2^* & -c\,x_2\,x_3^* \cr
-c\,x_3\,x_1^* & -c\,x_3\,x_2^* & 1-c\,x_3\,x_3^*
\end{pmatrix}
\eea
with
\beq
c=\frac{1}{\sqrt{1+|x|^2}\left(\sqrt{1+|x|^2}+1\right)}
\eeq

The down quark and charged lepton mass matrices of Eq. (\ref{lightmass}) can be diagonalized readily.
Their eigenvalues are given by:
\begin{eqnarray}
&~&m_1^2+m_2^2+m_3^2 = \frac{|d_1|^2(1+|x_2|^2+|x_3|^2)+|d_2|^2(1+|x_3|^2+|x_1|^2)+|d_3|^2(1+|x_1|^2+|x_2|^2)}{1+|x|^2} \nonumber \\
&~&m_1^2m_2^2+m_1^2m_3^2+m_2^2m_3^2 = \frac{|d_1|^2|d_2|^2(1+|x_3|^2)+|d_2|^2|d_3|^2(1+|x_1|^2)+|d_3|^2|d_1|^2(1+|x_2|^2)}{1+|x|^2} \nonumber \\
\label{masses}
&~&m_1^2m_2^2m_3^2 = \frac{|d_1|^2 |d_2|^2 |d_3|^2 }{1+|x|^2},
\end{eqnarray}
where $d_i$'s are common for $M_D$ and $M_E$, while the $x_i$'s are different.  From Eq. (\ref{masses}), it follows that realizing the
mass hierarchy is possible only when $|d_i|$ are hierarchical, $|d_1| \ll |d_2| \ll |d_3|$, in which case we can write down very simple formulas for the three masses:
\begin{equation}
m_i = |d_i| \cos\theta_i~.
\end{equation}
Here we define three mixing angles as:
\begin{equation}
\tan\theta_1 = |x_1|,~~\tan\theta_2 = \frac{|x_2|}{\sqrt{1+|x_1|^2}},~~\tan\theta_3 = \frac{|x_3|}{\sqrt{1+|x_1|^2+|x_2|^2}}
\end{equation}
with $0 \leq \theta_i \leq \pi/2$.  These are the same definitions used in Eq. (\ref{thetas}).

Noting that the mass matrix elements of Eq. (\ref{initial}) can be all made real by redefinitions of fields,
we also obtain the unitary matrices that diagonalize $M^D$ and $M^E$:
\bea
U^T M^D V &=& M^D_{diag}\\
V^T M^E U &=& M^E_{diag}
\eea
We interchanged the notation $U\leftrightarrow V$ passing from $D$ to $E$, because it is $M_E^T$ that has the
same form as $M_D$. Again, the matrices $U,V$ are different for down type quarks and charged leptons, we use
the same symbol however.  The unitary matrices $U$ and $V$ are given as (with $|d_1| \ll |d_2| \ll |d_3|$
\begin{eqnarray}
\label{U}
U \simeq \left(\begin{matrix}1& -\frac{m_1}{m_2} t_1 s_2 & -\frac{m_1}{m_3} t_1 c_2 s_3 \cr \frac{m_1}{m_2}t_1 s_2 & 1 & -\frac{m_2}{m_3}t_2 s_3 \cr
\frac{m_1}{m_3}\frac{t_1 s_3}{c_2} & \frac{m_2}{m_3}t_2 s_3 & 1\end{matrix}\right) ~,
\end{eqnarray}
\begin{eqnarray}
\label{V}
V \simeq \frac{1}{1+c_1c_2c_3} \left(\begin{matrix}c_1+c_2 c_3 & -s_1s_2 & -s_1c_2s_3 \cr s_1 s_2 c_3 & c_2+c_3 c_1 & -s_2 s_3 \cr s_1 s_3 & c_1 s_2 s_3 & c_3 + c_1 c_2\end{matrix}\right)~.
\label{V}
\end{eqnarray}
Here $c_i = \cos\theta_i,~s_i = \sin\theta_i,~ t_i = \tan\theta_i$. Terms of order $(m_2^2/m_3^2)$  and $(m_1^2/m_2^2)$ are ignored in the derivation of these matrices.

It is possible to fit all quark and lepton masses consistently to the observed values.  The mixing angles are related by the ratios
given in Eq. (\ref{massangle}).

\end{document}